\documentclass{camera}
\usepackage{graphicx}  
\begin{document}

%
\title{GRBs with the Swift satellite}

%
\author{Guido Cincarini(1,2) on behalf of the Swift Team}

%
\organization{

(1)Universita Bicocca, Piazza della Scienza 3, 20126 Milano

(2)INAF - Osseravatorio Astronomico di Brera, v. E. Bianchi, 46, 23807, Merate (LC)

}

\maketitle

\begin{abstract}
I touch upon some of the discoveries made by the Swift Team during the
first 18 months of operation focusing on a few critical points. In
addition to the early afterglows and complete coverage of the light
curves, we mention the discovery of the location of Short bursts, the
statistics on flares that is under completion and refer to their
explanation as due to late internal shocks. The full understanding of the
connection with supernovae and the study of high z flares will
ultimately lead to the detailed understanding of the mechanism of
explosion both for GRBs and for Supernovae.

\end{abstract}

%

\section{Introduction}

Following the launch of the Swift satellite, Gehrels et al. (2005) the
Swift Team, and the community at large, witnessed an unprecedented
collection of data. The instrument performed excellently and more than
anything else many of us felt that the key advantage consisted in the
rapidity by which the spacecraft was able to point the target on the
axis of the Narrow Field Instruments (NFI) and in a perfect worldwide
organization that allowed the real time data analysis and follow up
both from the ground and from space. In addition the performance of
the spacecraft, the ASI (Agenzia Spaziale Italiana) ground station was
far beyond specifications thanks to the Malindi staff so that all
the data were, and are being, received quickly and without
losses. Now, at about 18 months from launch, we have a reasonable
understanding of the data and we can tackle the statistical studies
since the sample is large enough (we detect about 102 bursts per year)
to allow it. It is also time to improve theory and interpretation. The
data are now good enough that in spite of the fact that the fireball
model (Rees \& Meszaros, 1992) has been confirmed in its more general
features, many details are lacking and such details, once they
are related to the observations, are fundamental to understand the
physics at work. The work carried out with Swift has been, so far, a
Team work where everybody participated to the research we carried out
in different roles and capabilities. The input of the theoretician has
been extremely useful since thanks to them the team was greatly
facilitated in the understanding of the models and of the observations
and move swiftly since the very beginning.

It is an impossible task to evidence in a few pages the richness of
the data we obtained in this period, data that relate to a large
number of bursts and involve other satellites and the largest ground
based facilities of the world. In addition very many review articles
and excellent journal articles are being published in the field. We
will simply try to touch upon the most important issues and briefly
illustrate some of the data we obtained.  In the following we will
discuss some of the highlights of the Swift observations and
discoveries.  In the second section we describe the theoretical
framework.  In the third section we will discuss the morphology of the
light curves.  To give a broader understanding of the capability of
the Swift mission and to illustrate the power of combining
observations by different satellites, in Section 4 we briefly discuss
the burst GRB060124.  In the Section 5 we will deal with flares. One
of the major discoveries carried out by Swift has been the location of
the short GRBs. This will be discussed in Section 6. Finally we will
touch upon two fundamental issues again related to discoveries made by
Swift, the connection between GRBs and SNe, Section 7, and the
detection of high z objects, Section 8. We conclude with the host
galaxies and progenitors with the conclusion that the real research
work and deep understanding of what is going on just started.

\section {Generalities on the model}

The framework of the discussion in this paper is that of the
collapsar (Woosley 1993, MacFadyen \& Woosley 1999,
Zhang \& Woosley, 2003) while for the emission we refer to
the fireball model by Rees \& Meszaros (1992) and following
modifications. 
In the collapse of the iron core of a massive rotating star we
likely end up with a black hole and an accreting disk. In the process
a large amount of energy is produced that finds its way out the star
as a well collimated jet, $\sim$10 degrees, and later impact in the
interstellar medium. The decrease of the external pressure after the
break out causes the front of the jet to expand. The central engine,
black hole and accretion disk, has a scale length of about 10$^6$cm
and the energy released is of the order of 10$^{50}$ -- 10$^{52}$ erg. The
emission is turbulent and relativistic pellets move at different
Lorentz factors causing an internal shock when a faster pellet (shell)
catches up with a slower pellet. The physics of the encounter between
two pellets, conversion of kinetic energy in internal energy and
emission of electromagnetic waves, is rather straightforward (see for
instance Kobayashi et al., 1997). The scenario becomes now somewhat
more complex. In the immediate vicinities of the progenitor we
certainly have a small ionized region of the interstellar medium or a
cocoon formed by highly absorbing material. Wind, as observed in
massive stars and Wolf Rayet stars, may be present. It is important to
disentangle the various components of the environment in order to
better understand the physics, the characteristics of the environment
and of the progenitor. A high
Lorentz factor ($\gamma$) pellet may be expelled first and this could
interact with the ISM before the internal shock occurs. While detailed
work and simulations are going on in order to fully clarify the
evolution of the ejected material, it suffices to remind now that in
addition to the clashing pellets (shells) causing the so called
internal shock we have an external shock as soon as the material
encounters the interstellar medium. Shocks, as it is well known, are
characterized by a forward and reverse shock with Synchrotron and
Inverse Compton emission. It is striking that such a complex set of
phenomena will originate, as we will see in the following, a quite
simple observed output with rather standard characteristics.  As we
know, and we will describe later, Swift located also the short
bursts. The short duration of the prompt emission rules out
immediately the collapse of a massive star for which the free fall
time is too long ,t$_{ff} \propto M^{-1/2} R^{-3/2}$. These seem to be due to
the merging of a NS-NS or NS-BH binary and the events show characteristics
in the light curve that are very similar to those observed in the long bursts.
The basic physical phenomena must be simple and similar.

\section{Morphology of the light curves and steep early decay}
\begin{figure}
\includegraphics[width=\columnwidth]{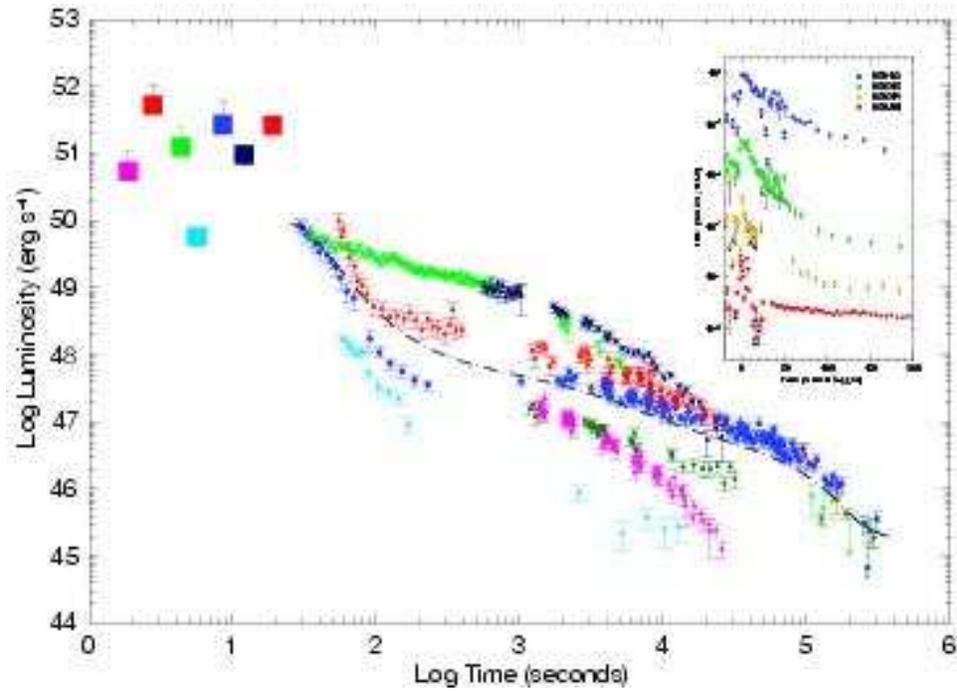}
\caption{ Light curves of GRBs non presenting Flares in the source rest frame. 
The filled squares refer to the mean BAT Luminosity. The dashed dotted line is the mean curve. 
Note the mild decay of GRB050401. In the inset we plotted both the BAT (transformed to the XRT pass-band ) 
and XRT data in order to evidence the continuity between the BAT observations and the XRT observations.}
\label{fig01} 
\end{figure}
\begin{figure}
\includegraphics[width=\columnwidth]{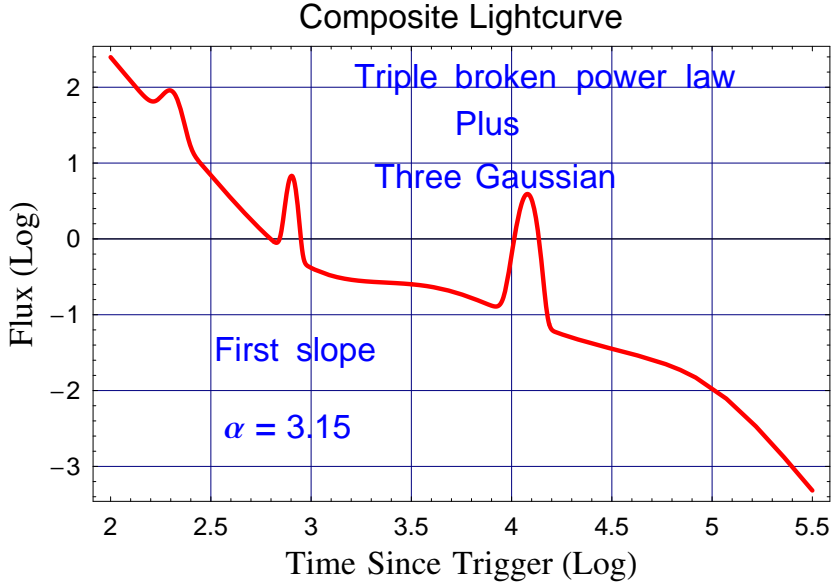}
\caption{This composite light curve, without the flares however, reflects a possible fitting of the light decay 
of GRB050814. The curve is fit by a triple broken light curve with
different slopes. In this case the second mild slope is slightly
increasing in brightness as it happens in various cases. There is 
somewhat of a degeneracy between this model and that of a broad bright
flare of low intensity. In both cases we need injection of energy.}
\label{fig02} 
\end{figure}
\begin{figure}
\includegraphics[width=\columnwidth]{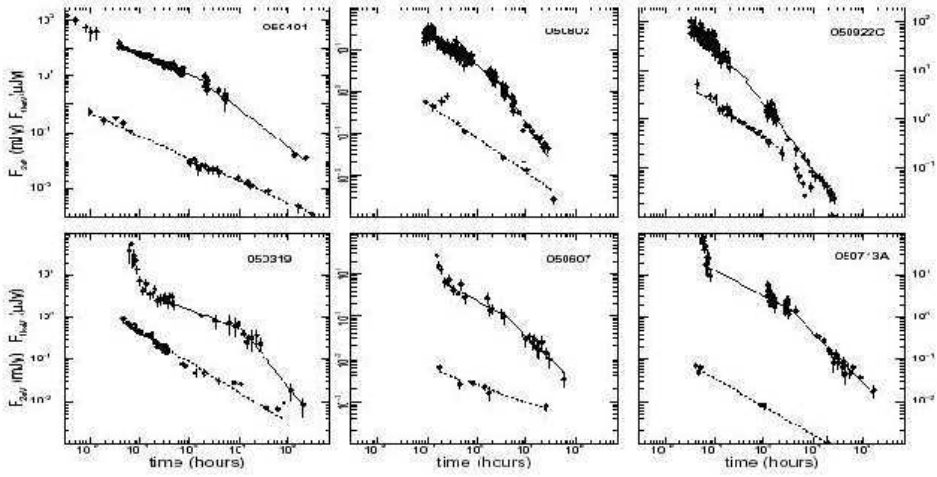}
\caption{Some of the cases in which the X ray break is not observed at
optical frequencies (Courtesy of Panaitescu).}
\label{fig03} 
\end{figure}

The basic features of the X-ray light curve were discussed in
Chincarini et al. (2005) and later by Nousek et al. (2006),
O' Brien et al. (2006). It just happened that
the first set of light curves observed by XRT and for which we had an
estimate of the redshift did not show any flare, see Section 5. These
curves, however showed the whole story. 
We had continuity
between the BAT light curve and the XRT light curve (see the inset in Fig.~\ref{fig01}
from Chincarini et al. 2005). They showed a steep decay at the very
beginning of the XRT light curve, Fig.~\ref{fig01} and Tagliaferri et al. (2005), 
except in some cases, as GRB 050401, where the initial decay
is rather smooth, and showed the presence of 1 to 4 breaks. The
morphology of the light curves in other words is not the same for all
the long (T$_{90}$ $>$ 2s) GRBs. The cartoon, Fig.~\ref{fig02}, shows the various
morphologies we might have. 
 The steep early decay, starting at the
onset of the XRT observations, may be due in most cases to the tail of
the prompt emission and related flare activity. Such model would
easily explain the steep decay since by measuring it from the
beginning of the flare (see next section) and eventually normalizing
it to the rising time of the flare prompt emission it would be within
the limits posed by the curvature effect where $\alpha$ = 2+$\beta$
with Flux $\propto t^{-\alpha} \nu^{-\beta}$ (Kumar \& Painatescu 2000).
The continuity between the BAT prompt emission and the XRT
emission supports the model. More problematic is the modeling of the
second phase of the light curve where we observe a mild decline
$<\alpha>$ $\sim$ 0.7. To stop the steep decay of the light curve we need
to inject energy. What is striking is that not only we need energy
injection but in most cases this needs to be done smoothly since the
light curves do not generally show, during this phase, strong
oscillations. Oscillations or fluctuations are observed in some cases
however. Panaitescu (2006) discusses he energy injection model following the 
steep decay, on the other
hand it seems to us we really need too finely-tuned injection
to have such smooth light curves.
Finally we have a series of breaks. Late
breaks with steepening of the light curve decay have been interpreted as
mark of the time when the relativistic beaming angle of the eject
equals the jet opening angle. Such a break should be achromatic since
it is simply due to geometrical factors. Here our knowledge is rather
confused. The number of achromatic breaks that have been observed is
rather small. Often we did not observe a break when expected and in many
cases the break observed in the X-ray band does not match the break
observed in the optical, Fig.~\ref{fig03}, (Panaitescu 2006b). 
The latter lack of coincidence may
be due to a different origin of the breaks in the two bands and in any
case these uncertainties put a shadow on the use of the break to
estimate the jet opening angle and emitted energy. This and the lack
of an accurate control on the bias of the samples used, generally flux
limited samples, make very uncertain, even if obviously promising due
to their detection at large cosmological distances, the use of GRB
sample to estimate the cosmological parameters (Ghirlanda et al. 2004).
The energy, or Fluence, involved in the XRT light curves is likely a
mixture of late prompt emission and afterglow, in some cases it is
similar to the energy observed during the prompt emission.
\begin{figure}
\includegraphics[width=\columnwidth]{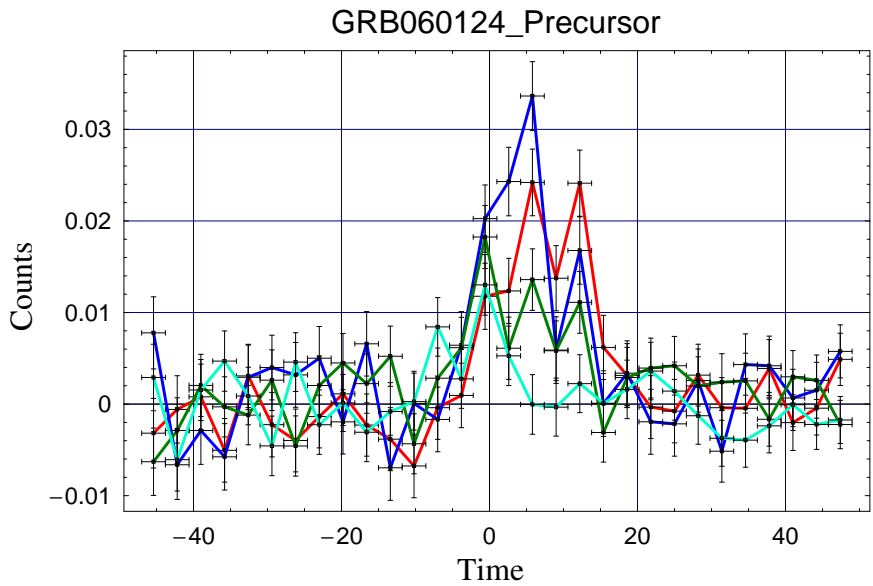}
\caption{The precursor of the GRB0670124. The total fluence of the precursor is 
1.24 10$^{-6}$ erg cm$^{-2}$ while the fluence of the prompt emission
is 2 10$^{-5}$ for the prompt emission at t = 550 s and 2.93 10$^{-6}$
for the BAT emission at t = 700 s (for the shape of the prompt
emission see Figure 1 in Romano et al. (2006b).  The precursor is about
4.1\% of the prompt emission (for uniformity to other GRBs we did not
add the emission in the band 0.2-10 keV, XRT).}
\label{fig04}
\end{figure}

\section{GRB060124. Precursor and Multi-wavelenght observations}
GRB060124 (Romano et al. 2006b and references therein) is a beautiful
example of multi-wavelength observations and of a complete temporal
coverage on a very broad range of frequencies. BAT was triggered on
the precursor so that the narrow field instruments were already on the
source at the time the main burst occurred. It is located at a
distance z = 2.297. While we refer the reader to the paper above for a
full discussion, here we will only touch upon a few details. The
precursor precedes the GRB by about 500 s, it is rather soft and in
about 20 s emits a few percent of the prompt emission energy, that is
an energy of the order of 10$^{49}$ - 10$^{50}$. The emission starts
off rising in a similar way in all channels and however it dies off
after a few seconds in the 100-350 keV channel while it lasts for
about 20 seconds in the softer BAT channels, Fig.~\ref{fig04}. What is
a precursor telling us? We must take into account this time scale in
the modeling. This needs to be looked upon carefully since not only it
is the messenger of the GRB but may also carry information on how the
event is going to perform.
\begin{figure}
\includegraphics[width=\columnwidth]{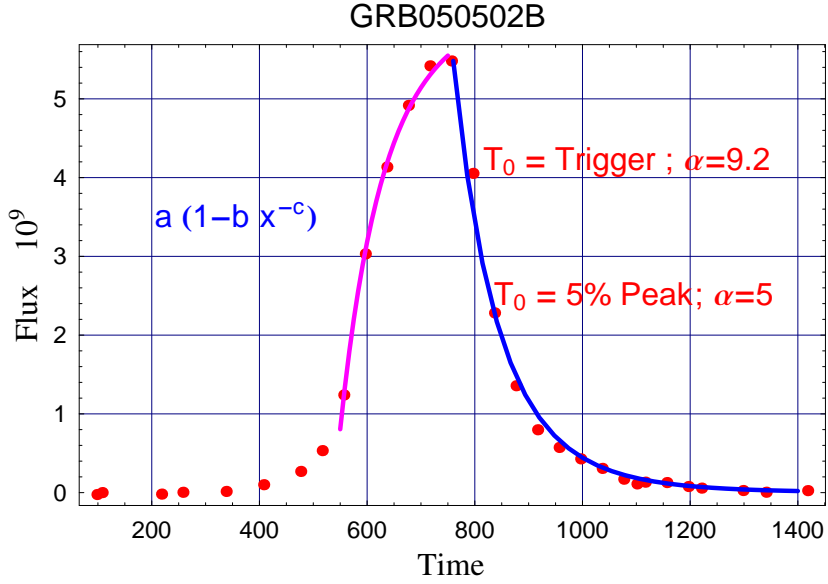}
\caption{This is one of the brightest and cleanest (no blends) flares observed. 
The slope of the decay is larger than the limit imposed by the naked
burst unless the slope of the power law is computed with T0 set at
the maximum of the flare. It also fit the normalized profile
suggested by Kobayashi for the prompt emission.}
\label{fig05} 
\end{figure}
The burst is very energetic and the prompt emission in 510 s emits
E$_{iso} (1-10^{4}$ keV) = 4.2 10$^{53}$ erg. The prompt emission has been observed
both from the BAT and XRT instrument. It is noisy and presents a
morphology slightly different from normal flares (see next section)
but the XRT curve has similarities albeit very high luminosity
(compare to the flare in GRB050502B however). The hardness ratio of
the prompt emission follows almost in phase the flaring of the prompt
emission and the spectral index soften passing from the first flare to
the second. A behavior that we will also find in the flares we are
dealing with in the next section.

\section{Flares}
Early in the mission we observed two unusual events GRB050406 (Romano et al. 2006a)
and GRB050502B (Falcone et al. 2006a), Fig.~\ref{fig05}. Similar events were
also detected in some of the GRBs observed by Beppo SAX and analyzed
only later (Piro et al. 2005). The XRT light curve gives right away the
impression it consists of two components: a standard component whose
morphology follows the shape and details discussed in section 2, with
superimposed a more or less large and intense bump that we call
flare. Such morphology has implications on the physical model.  The
two curves are unrelated and generated by different collisions. The
running of time for the flare, therefore, is not related to the
running of time dictated by the beginning of the prompt emission. In
other words the flare has its own clock. Indeed an interesting
analysis carried out by Liang et al. (2006) shows that, assuming that
the decay of the flare can not violate the curvature slope, $\alpha$=
2+$\beta$, the origin of time T0 from which to compute the evolution
of the event is near the beginning of the flare itself. 

\begin{figure}
\includegraphics[width=\columnwidth]{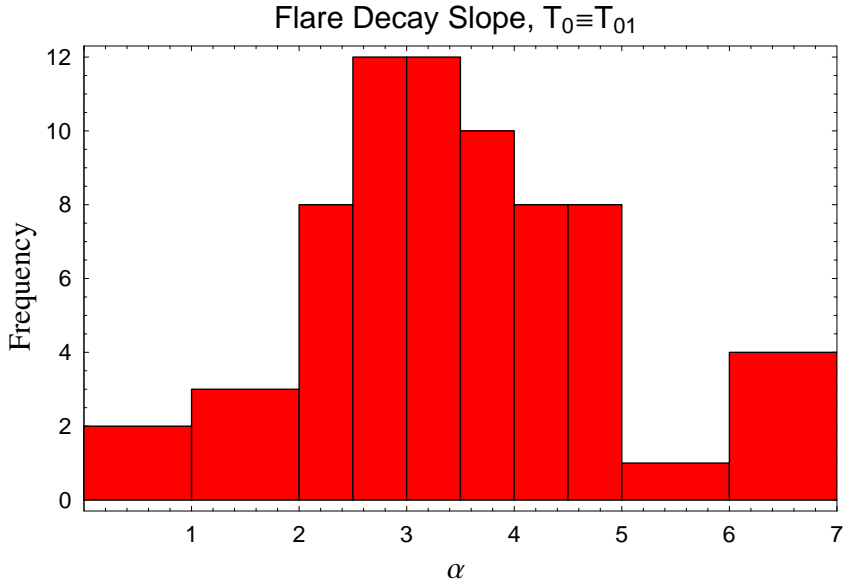}
\caption{Distribution of the decay slopes computed using as T0 the point where the flux 1\%
of the peak (T$_{99}$). This procedure is self consistent but overestimates
somewhat the slope due to the early T0. For the largest slopes see
GRB050502B, Figure 5.}
\label{fig06} 
\end{figure}

The morphology of flares should reflect the characteristics of the
shock generating them. To isolate them from the composite light curve,
in Chincarini et al. (2006a) we subtracted from the observations the
standard light curve after fitting it with a broken power law. The flares are
characterized by various morphologies. The cleanest shape is that
suggested by the flare in GRB050502B. This fits very well the
Kobayashi profile or, equally well, an exponential profile of the form
f$_{\rm rising}$=a(1-e$^{-b(x-c)}$), followed by a power law decay,
see also Dermer (2004,2006).

In some of the flares the rising part is better fit by a simple power
law. On the other hand, flares are very complex and are not isolated
events (see, for instance GRB051117A). In other words they are due to
the collision of a set of pellets, very similar to the activity we
have during the prompt emission. Quite often, therefore, as it may be
the case for the simple power law fit of the rising part of the curve,
the true profile may be masked by multiple flares, low count rate and
sampling. The decay, as stated above, is generally well fit by a power law,
Fig.~\ref{fig06}.
\begin{figure}
\includegraphics[width=\columnwidth]{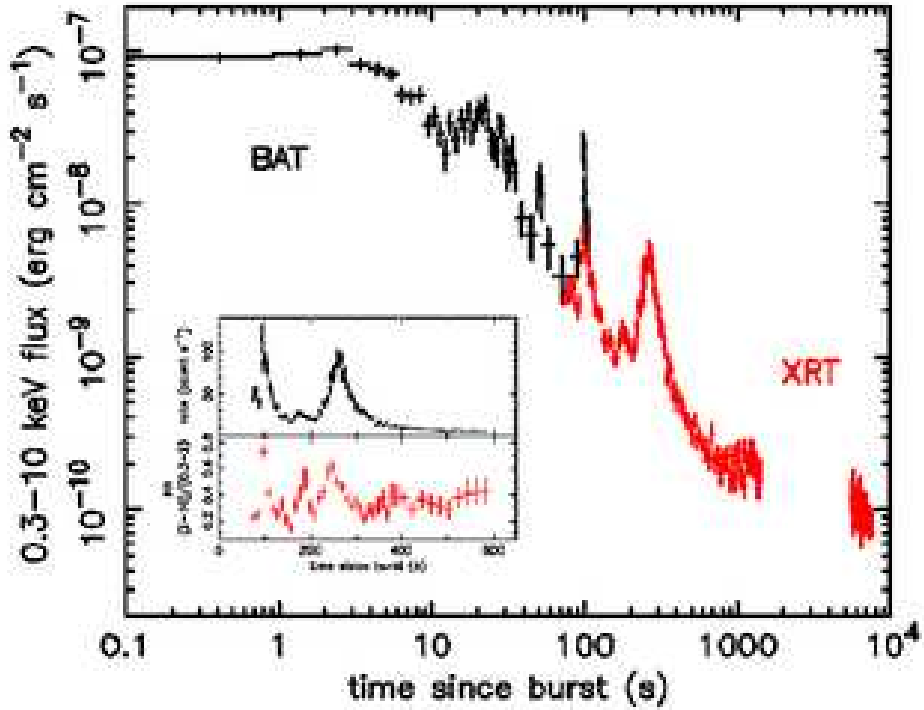}
\caption{GRB060607: this is one of those GRBs that tell the whole
story. We notice continuity between the BAT and XRT light curve. The
early flare, related to the prompt emission, are detected both with
BAT and with XRT. The hardness ratio follows the light curve of the
flare almost in phase hardening in its rise to the maximum and
softening afterwards see inset.}
\label{fig07} 
\end{figure}
The spectrum and the hardness ratio variations during the flares show
some resemblance with the behavior of the spectrum characteristics of
the prompt emission (Falcone et al. 2006a). In the case of GRB050502B,
for instance, a Band function or cut--off power law fits better than
a simple power law and spectral evolution is evident. In some cases the
use of a black body + power law is requested (Falcone et al. 2006). The
hardness ratio shows quite clearly that the
spectral evolution is almost in phase with the flare light curve
hardening during the rise to the peak to soften again in the decaying
phase, Fig.~\ref{fig07} (compare also with GRB060124).
\begin{figure}
\includegraphics[width=\columnwidth]{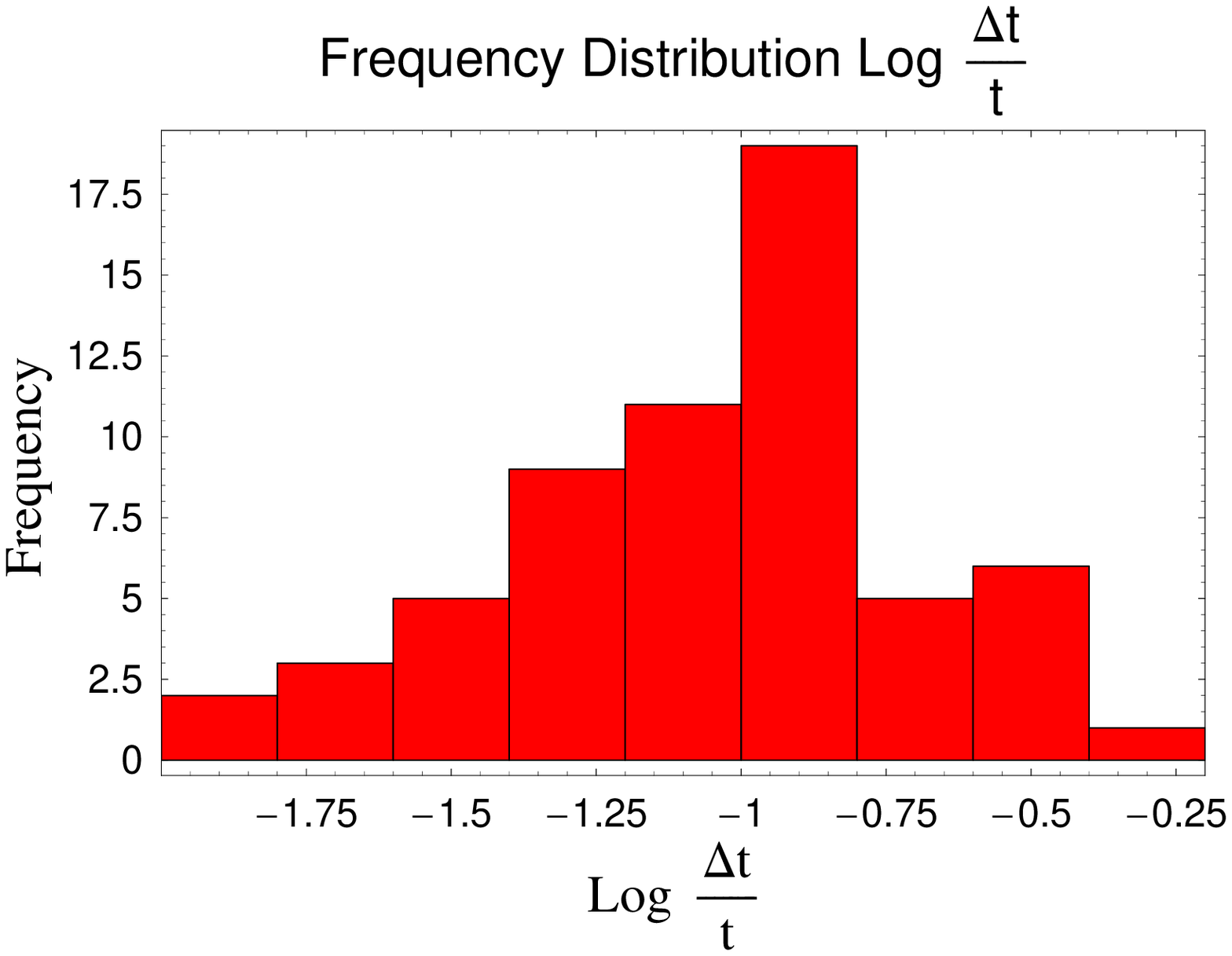}
\caption{Frequency distribution of $\Delta$t/t where $\Delta$t is given by the sigma of the Gaussian fit. }
\label{fig08} 
\end{figure}
A sample of 61 Flares has been analyzed by Chincarini et al. (2006b),
Falcone et al.(2006) to measure statistical parameters as $\Delta$t/t
(flare duration versus time of occurrence), decay slope, decay to rise
ratio, flare energetic, flare to burst fluence and spectral
parameters. In addition simulation were carried out to understand the
selection effects and biases introduced by the mode of observation and
sampling characteristics especially at later times and low
fluxes. Here we show only the $\Delta$t/t and refer to Chincarini
2006a, or to Chincarini 2006b, for further details. To have a robust
and unbiased statistics we plotted the results shown in Fig.~\ref{fig08} where
the estimate of the flare duration is given by the sigma of the
Gaussian (T$_{\rm sigma}$ used in the fit of the composite light curve). This
is a robust indicator and simulations show that it is not much
affected by the morphological model of the flare. On the other hand we
must keep in mind that the Gaussian fit has neither physical meaning
nor we can transform the T$_{\rm sigma}$ in T$_{90}$ easily. The statistics tell us
that the Half Peak Width (HPW) distribution of the flare peaks at
about 3/10 of the time of occurrence. On the other hand it is
customary to estimate the parameter $\Delta$t/t, especially in the
theoretical modeling, with $\Delta$t = T$_{90}$. To use this parameter we must
perform a more detailed fit and use only the flares for which we have
a reasonable good statistics. The distribution of T$_{90}$/t (where T$_{90}$ has
been derived in this way) is given in Fig.~\ref{fig09} with the peak of the
distribution at about 5/10.  These figures are in agreement both with
the internal shock model as given by a continuously active engine or
by internal shocks due to the lazy shells model (refreshing), see
discussion below.
\begin{figure}
\includegraphics[width=\columnwidth]{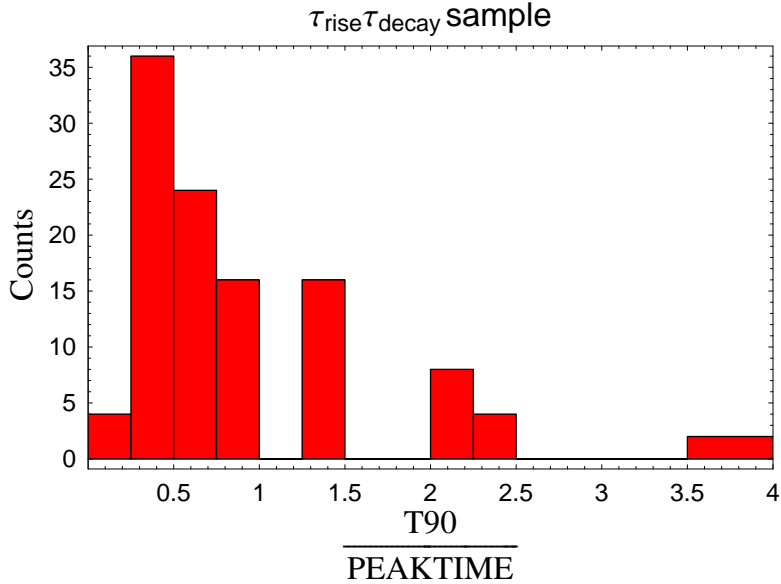}
\caption{The distribution $\Delta$t/t with $\Delta$t = T$_{90}$. The
large values are due mainly to blends of flares and will be discussed
elsewhere.}
\label{fig09} 
\end{figure}

\section{Short Bursts}
\begin{figure}
\includegraphics[width=\columnwidth]{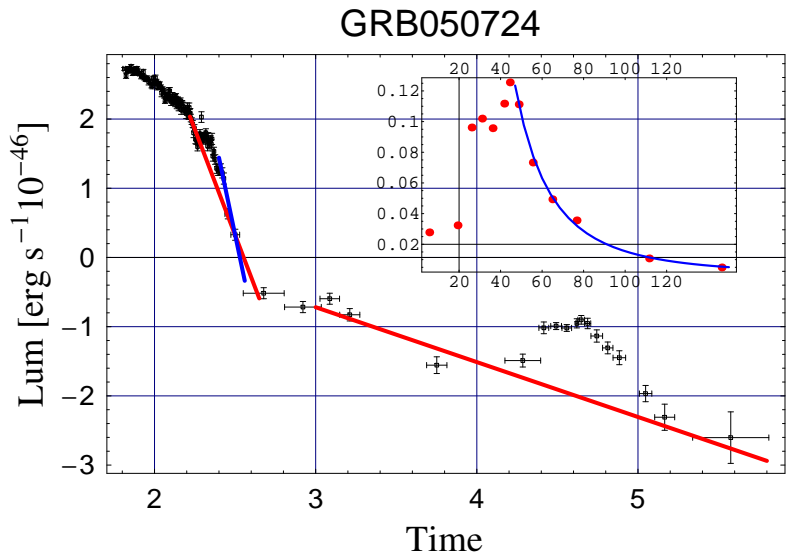}
\caption{The continuous lines illustrate the approximation by which the flare profile is isolated 
from the standard underlying light curve. It also evidences the
composite nature (blend) of the late flare. Finally the inset
evidences an early flare after subtraction of the main curve and the
excellent fit of the decay with a power law.}
\label{fig10} 
\end{figure}

One of the big discoveries made by Swift is the location of the short
bursts and in particular the location of GRB050509 next to an
Elliptical galaxy (Gehrels et al. 2005). This was the observation that
immediately guided to the NS-NS, NS-BH model. In one case, GRB050709
(Hjorth et al 2005, Covino et al. 2005), we observed a host galaxy of
irregular morphology that however showed a EA (Emission-Absorption)
line spectrum. That is we recognized the presence of dominant type A
stellar population indicating the existence, in addition to the gas
and HII regions, of an old population, old enough to let binaries
evolve and form relativistic binaries. We have detected so far
(updated to GRB060801 and including HETE detections) 18 shorts. For 6
of these we also have an optical detection and six redshifts, two of
which rather uncertain however. GRB050724, Fig.~\ref{fig10}, is one
of the most significant light curves we have so far. The X-ray light
curve has all the characteristics presented by long bursts showing
continuity between the BAT emission and the XRT emission, early
mini-flares and late large flare. The presence of flares is indeed a
fundamental point since it would indicate that such phenomenon is
completely independent from the environment since it occurs both in
star forming galaxies and in old population galaxies where we have
little gas and dust.
This would call immediately for an internal shock origin of the flares, late activity
of the central engine or lazy shells (refreshing). The inconvenient is
that this burst is on the border line of the distribution and we may doubt somewhat that
it is really a short (or do we have different types of short bursts?).
The initial spike lasted 250 ms followed
after about 30 s, as in GRB050709, by a soft bump that lasted
120s. The observed spectrum index was $\beta$ = 0.7 and the source was
detected in the radio, optical and near-IR.  The other open question
is whether the emission in this case is spherical or, as expected due
to the preferential axis defined by the BH-accretion disk model,
confined to a jet. In this case we need to estimate the opening angle.
Do we observe an
achromatic break in any of the observed light curves? A break in the
X-ray light curve, with XRT and Chandra, has been observed in GRB051221,
(Burrows et al. 2006). The estimated jet opening angle is
4$^{\circ}$ -- 8$^{\circ}$. We do not have any
way to know whether the break is achromatic. Except for a very few
uncertain cases all the detected shorts are at low $z$. The energy E$_{iso}$
is about 10$^{48}$ --10$^{49}$ erg, about a factor 100--1000 lower that
the energy involved in long GRBs. In conclusion, while in the field we
reached fundamental knowledge, we are far from having a complete data
set to carefully guide the interpretation and the theory. The space-based
observations and ground-based observations are hard to come and shorts
remain, and may remain, rather elusive. Finally are short burst giant
magnetar ?
Energy, spectrum and duration would be
in good agreement; magnetars, furthermore, manifest themselves more
than once so that monitoring (but the task is close to impossible) 
may pay off eventually.
\begin{figure}
\includegraphics[width=\columnwidth]{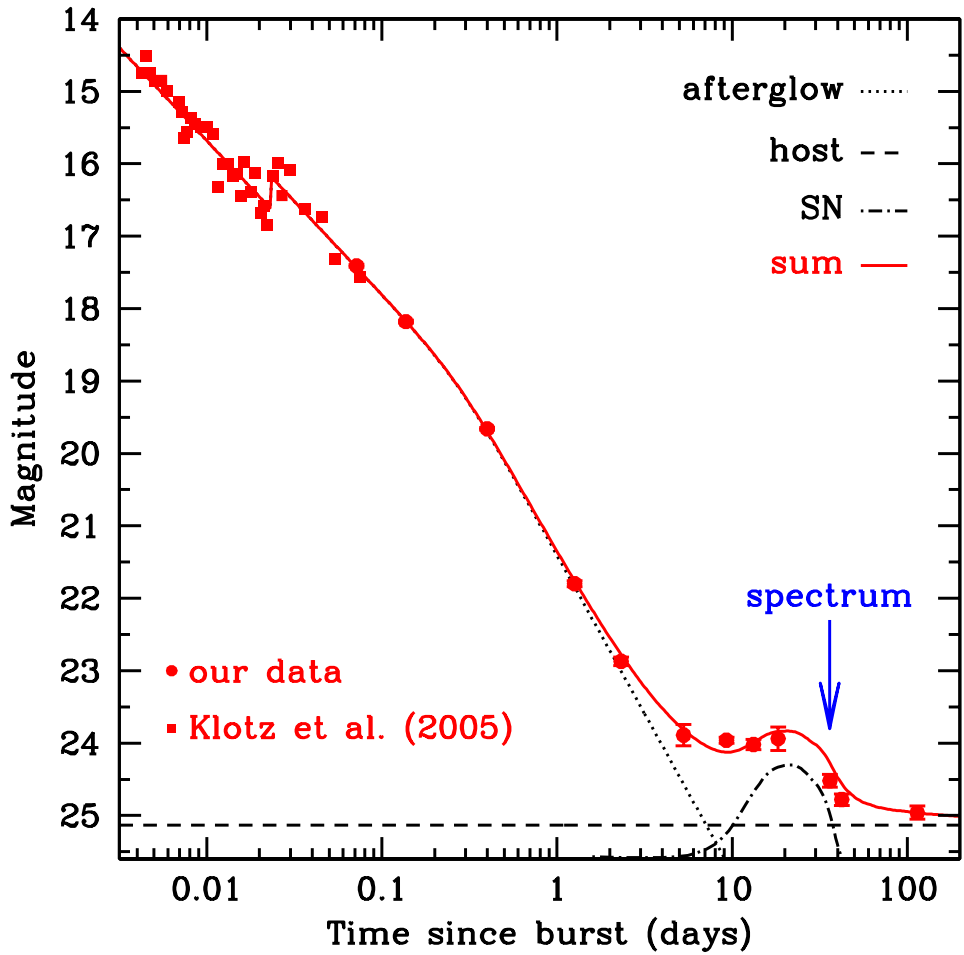}
\caption{The figure shows the light curve of GRB050525A where the classical bump indicates the presence of a Supernova.}
\label{fig11} 
\end{figure}

\section{GRBs and Supernovae}
We observed 5 golden cases demonstrating the connection of Supernovae
with GRBs. GRB980425, Galama et al. (1998), GRB 030329, Hjorth et
al. (2003), Stanek et al. (2003), GRB031203, Malesani et al. (2004),
Thomsen et al.(2004), GRB050525A, Della Valle et al.(2006), GRB060218,
Campana et al.(2006). The clear indication of the possible presence of
the Supernova, when the source is bright enough to be detectable over
the decaying curve of the GRB, is the presence of a bump, as shown in
Fig.~\ref{fig11} for GRB050525A, that occur generally 10 to 20 days from the
GRB trigger. The spectrum of the Supernova is obtained by subtraction
of the galaxy and it has been shown that, in all the cases, we are dealing with SNe - Ibc
Supernovae, rather luminous, M$_{V}$ $\sim -19$ and expansion velocities of the
ejecta up to v $\sim$ 30000 km/s. GRB-SNe are comparable at optical
frequencies to locally observed SNe. Radio emission is observed at
later time when the jet, impacting the interstellar medium,
decelerates to sub-relativistic speed and its emission becomes nearly
isotropic. The radio luminosities of the local events are about 104
times fainter than those of typical GRB-SNe. There
is, therefore, a clear distinction in the radio between the two types
likely due to the large amount of energy that the GRB ejecta deposit
in the interstellar medium, Soderberg (2006).
\begin{figure}
\includegraphics[width=\columnwidth]{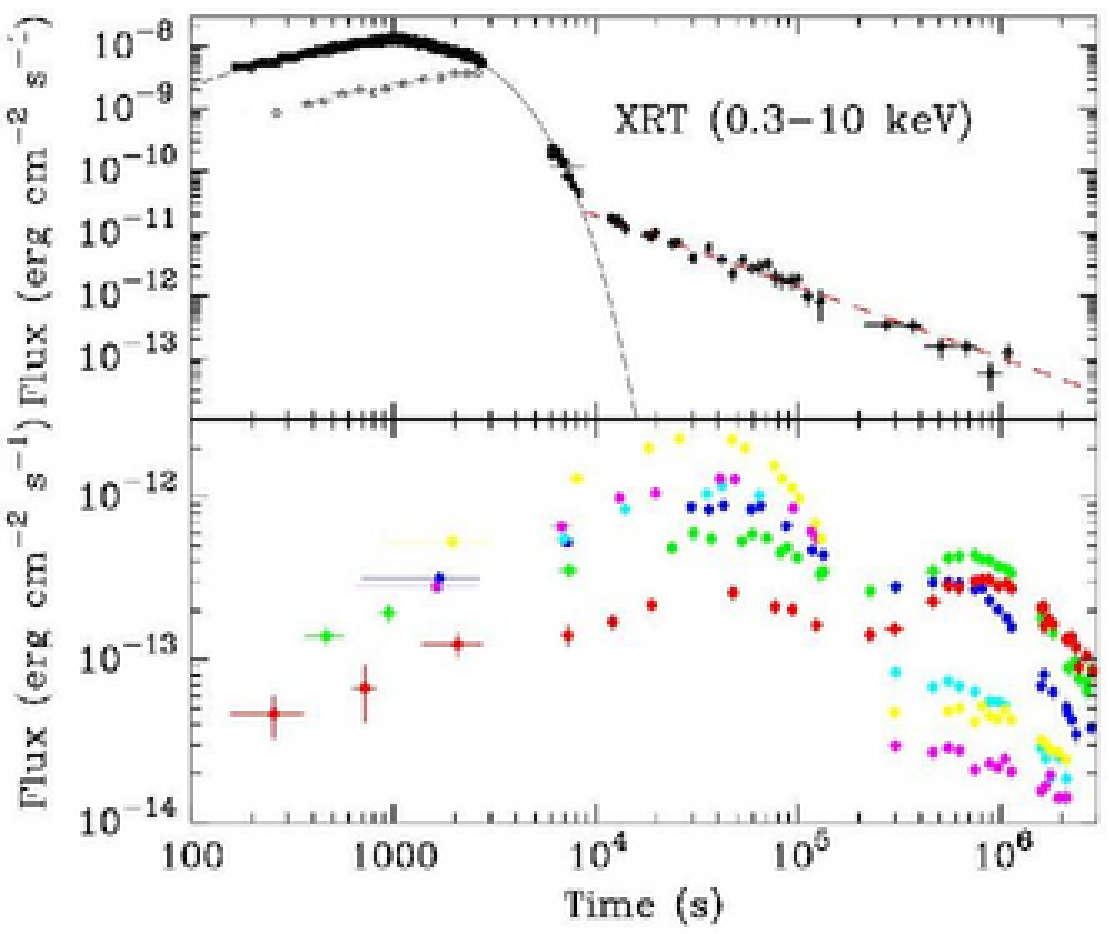}
\caption{The XRT light curve (upper panel) and optical light curve (lower panel) of GRB060218. 
Top: The open circles refer to the contribution to the light curve of
the thermal component. Bottom: The light curve as derived from
UVOT. From top to bottom the following filters: V, B, U, UW1, UM1, UW2
(from Campana et al. 2006).}
\label{fig12} 
\end{figure}
The observations of GRB060218, Campana et al. (2006), gave us the
unique opportunity to observe the evolution of the phenomenon in all
of its details since the very onset of the SN phenomenon with the
detection of a thermal component in its X-ray spectrum. We are
catching a SN in the act of exploding. We witness the break out of a
shock driven by a mildly relativistic shell into the dense wind of the
progenitor. The long lasting prompt emission, T$_{90}$ = 2100 s, shows a
rather smooth behavior at all frequencies (XRT and UVOT observed
simultaneously most of the prompt emission since the NFI were pointing
the target after 150 seconds). The peak emission moved from high to low
energies (Fig.~\ref{fig12}) roughly as T$_{Peak}$ = k 10$^{-0.58 \nu}$ with
an optical re-brightening at about 800 ks from the trigger. The XRT
spectrum shows a blackbody component, kT $\sim$ 0.17 keV, visible for about
10000 s. The peaks observed with UVOT allow following the evolution of
the thermal spectrum component and deriving from the Luminosity the
radius of the emitting shell. The resulting model is that of a shock
propagating into the wind of the Wolf Rayet star progenitor.

We have discovered that the death of massive stars leads to the
collapse and explosions that manifest themselves either as GRBs or
SNe. In a few cases it has been possible however to observe both
events in the same source as in the cases listed above. The rate of
GRB-SNe is likely about 0.003 the rate of all massive star deaths,
(Woosley \& Heger, 2006), and observations should finally tell us
whether we have a SN (spanning however a large range of Luminosities)
in each GRB or whether the collapse of a massive star evolves
following different paths according to the value of parameters as
mass, angular momentum and metallicity. Whatever the answer may be the
fundamental point of the connection is that GRBs may serve as a
guideline to better understand the mechanism, and possibly solve the
long standing problem of the SN explosion, since we have additional
information related to the core collapse. For a detailed review on the
connection between GRBs and SNe see Woosley \& Bloom (2006).

\section{High z GRB, Host Galaxies and Cosmology}
\begin{figure}
\includegraphics[width=\columnwidth]{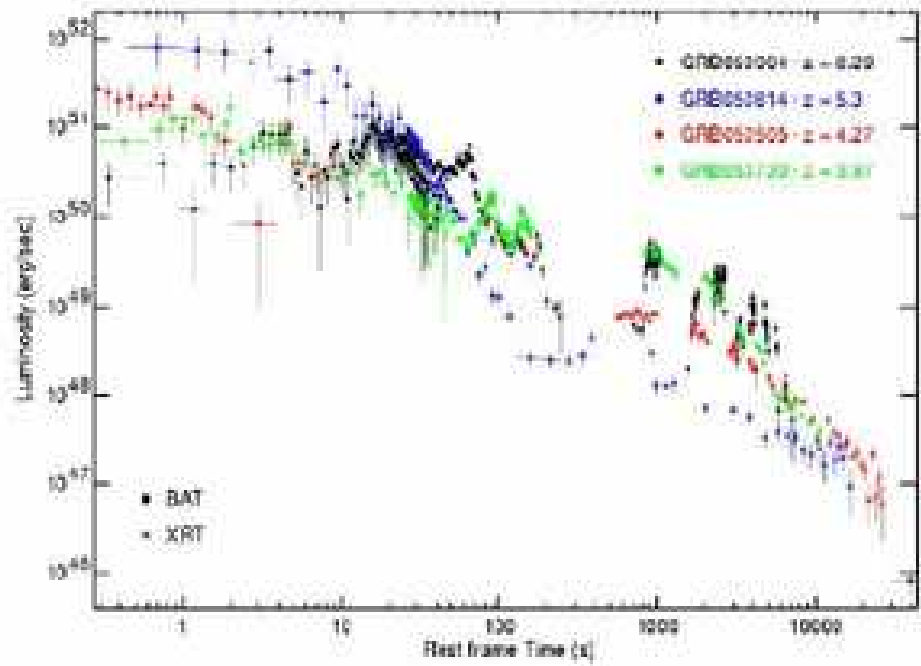}
\caption{GRB050904 and other high z GRBs observed by Swift. The early points are the BAT data converted to XRT. 
Note the very high flare activity and how the early flares are visible
with bith instruments. Flare activity is spread over most of the light
curve.}
\label{fig13} 
\end{figure}
It is well known that the redshift distribution of the bursts detected
by Swift is skewed toward high redshifts when compared to previous
detections. This means that we have the possibility, once we have a
reasonably large sample, to study under a new point of view the
intergalactic medium and host galaxy interstellar medium as a function
of redshift.  The radiation we observe carries information on the
environment surrounding the progenitor (Berger et al, 2005, detect
absorption lines generating in the wind of the proto-star), on the
interstellar medium of the Host Galaxies (HG) and on the IGM. The
estimate of the abundance may allow the tracking of the metal
production up to the re-ionization epoch. The large amount of
information we are getting and will get are easily understood since we
move to some extent along a track already sketched by the studies of
the AGN and we will complement that work adding new capabilities
especially thanks to the lack of the proximity effect. The highlight
of the Swift high z detection is GRB050904 at a redshift of z=6.29,
Cusumano et al. (2005), Tagliaferri et.al (2005), Kawai et
al.(2005). The composite picture we show in Fig.~\ref{fig13} is aimed
to evidence, among other things, the high frequency of bright flares
in high z objects. The frequency of activity, and of GRBs as a
function of the metal abundance, will be fundamental to the
understanding of the mechanism of the core collapse, whether the
choice to collapse into a GRB-SN or into a SN is dictated by rotation
influenced by the metal abundance of the progenitor and how much
governed by the laws of chaos.  Following Woosley \& Janka (2006) if
the star preserves high rotation velocity during its evolution
(essentially bypassing the red giant phase or losing the envelope to a
companion) the resulting Wolf-Rayet star will preserve angular
momentum assuming the metal abundance is low. Otherwise the rather
large winds (mass loss) will carry away also angular momentum and
compromise the progenitor in its path toward a GRB. Indeed rotation
and lack of symmetry are essential to produce a GRB (Zhang et al.,
2005). GRB050904 is obviously very luminous and may have involved in
the process a very massive star. The flare activity is very high,
larger at late time than in other bursts, and distributed practically
over the whole light curve so that Luminosity and metal abundance may
also be two parameters related to the flare activity and decreasing
opacity. The problem is now detection, one case is not enough for
statistics. Swift should insist on searching for these very rare
targets.  Furthermore for those GRBs without optical counterpart we
need to know accurate position in order to track down the host
galaxy. A step forward in this direction has been done by Moretti et
al. (2006), improving the typical XRT error circle from
6.5$^{\prime\prime}$ to 3.5$^{\prime\prime}$. A team effort lead by M.Goad
is underway to further improve the XRT astrometry.  Here we may get
fundamental knowledge in evolution and cosmology.  On the other hand
we are far from a good knowledge of the luminosity function that is
the base for statistical studies and the basic for the use of the GRBs
as cosmological probes. The most dangerous thing, as always in
astronomy, is the presence of unknown bias.  The least known
statistical bias is related to the probability of detecting an object
of a given Energy at a distance z. Assuming, for simplicity (easy
integration) and in order to clarify the concept, a Schechter
Luminosity Function in a survey limited in flux by the sensitivity of
the detector, (for simplicity we disregard pass-band and spectral
distribution of the source and assume for the selection function a
step function) the counts as a function of z are given by:

$$ N(>f)= \phi^{\star} \Gamma [\alpha+1,(E=D^2_L 4\pi f)/E^{\star}]~Vol~\Delta\Omega(f) $$

It is possible therefore to estimate in the plane redshift - Energy
the number of objects we expect, the minimum energy boundary for the
flux limited sample and the maximum energy boundary as determined by
the equation N(E,z) =1, that is the probability to find an object at
the Energy to be determined. With the present rate of detection and
redshift estimate we may be able to have soon a statistical
significant sample and attack some classical problems in Cosmology
having full control of the data.

\section{Conclusions}

There is no doubt that with the observations we obtained with the
Swift satellite we made, and are making, a fundamental step forward in
understanding observationally the evolution of the event and in
collecting data for statistical studies. At the same time it has been
possible to test theories and at least have an indication of what is
acceptable and what needs to be revised. Indications that came earlier
from observations from GRBs detected by Beppo SAX or other satellites,
were confirmed, as the GRBs-SNe connection, or denied, as the simple
power law decaying light curve, and the theory has now a more solid
ground on which to proceed. So far the interpretation of the data has
been done, by a large extent, on the existing theory that, in their
basic features, hold.  We showed that the field touches almost in all
branches of astronomy and is very demanding for the state of the art
instrumentation. Knowledge of the performance of the central engine
and of the progenitors is intimately related to the complex physics of
the death of massive stars and to all those phenomena accompanying the
evolution of relativistic stars. The evolution, and physics of the
collapse, is intimately connected to the GRBs - SNe connection and to
the abundance of the progenitors and of the host galaxies. This open
the field to evolution in relation to cosmology and at the end the
high luminosity and the possibility to be detected at very large
reshifts make them a fantastic beacon to probe the high z Universe and
the interstellar medium.  The most important observations seem to be
those capable of detecting high z GRBs that could be immediately
followed up by ground based observations with large and robotic
telescopes. Robotic telescopes are essential to ease such search. We
may need to think about desgning a few two meters aperture robotic
telescopes. While all the data tend to suggest we might have more long
bursts at high redshifts either because of the characteristics of
the progenitors and host galaxies or because of statistics based on
what we observed so far or simply because the phenomenon seems to
prefer low abundances galaxies it remain a puzzle why so
far we detected only one GRB at 6.29 (the detection of a possible
object at higher redshift, GRB060116 remains uncertain). How many did
we loose because of limitations related to the zone of the sky in
which the bursts have been detected and the difficulty of a proper and
immediate follows up? 

\section{Acknowledgements}
I thank the Swift Team members, OAB, ASDC, IASF--PA, PSU, GSFC, UL and
MSSL for useful discussion. In particular I acknowledge Alberto
Moretti and Patrizia Romano for helping with the data in various
occasions and with the manuscript.  The work in Italy is supported by
the ASI grant I/R/039/04 and science support is given by the Ministry
of the University and Research (PRIN), at Penn State by NASA contract
NASS-00136 and at the University of Leicester by PPARC.



%

\begin{thebibliography}{9}

\bibitem{}Berger, E., Kulkarni, S.R., Fox, D.B., 2005, ApJ, 634, 501
\bibitem{}Burrows, D.N., Grupe D., Capalbi M., 2006, ApJ, submitted, astro-ph/0604320
\bibitem{}Campana, S., Mangano, V., Blustin, A.J., Nature, in press, astro-ph/0603279  
\bibitem{}Chincarini, G., Moretti, A., Romano, P., et al. 2005, astro-ph/0506453
\bibitem{}Chincarini, G., 2006a conf. proc. of {\it SWIFT and GRBs: Unveiling the Relativistic Universe}, Venice
\bibitem{}Chincarini, G., et al. 2006b, in preparation
\bibitem{}Covino, S., Malesani, D., Israel, G.L.,  2006, A\&A 447 L5
\bibitem{}Cusumano, G., Mangano, V., Chincarini, G., 2006 Nature, 440, 164
\bibitem{}Della Valle, M., Malesani, D., Bloom, J.S., et al.,  2006, ApJ, 642, L103
\bibitem{}Dermer, C.D., 2004, ApJ, 614, 284
\bibitem{}Dermer, C.D., 2006, AIP, conf. proc. of {\it GRB in the Swift Era}
\bibitem{}Dermer, C.D., 2006, conf. proc. of  {\it SWIFT and GRBs: Unveiling the Relativistic Universe}, Venice
\bibitem{}Falcone, A. D., Burrows, D. N., Lazzati, D. et al., 2006a, ApJ, 641, 1010
\bibitem{}Falcone, A., et al. 2006b in preparation
\bibitem{}Galama, T.J., Vreeswijk, P.M., van Paradijs, J., et al. 1998, Nature, 395, 670
\bibitem{}Gehrels, N., Chincarini, G., Giommi, P., et al., 2004, Ap. J., 611, 1005
\bibitem{}Gehrels, N., Sarazin, C.L., O'Brien, P.T., et al., 2005 Nature 437, 851
\bibitem{}Ghirlanda, G., Ghisellini, G., Lazzati, D., Firmani, C., 2004 ApJ 613L 13
\bibitem{}Golenetskii et al., 2005, GCN Circ. 4197
\bibitem{}Hjorth, J., Sollerman, J., M\o{}ller, P., et al. 2003, Nature, 423, 847
\bibitem{}Hjorth, J., Watson D.,  Fynbo, J., et al. 2005, Nature 437, 859
\bibitem{}Kawai, N., Kosugi, G., Aoki, K., 2006, Nature, 440, 184 
\bibitem{}Kobayashi, S., Piran, T., Sari, R. 1997, 1997, ApJ, 490, 92 
\bibitem{}Kumar, P., Panaitescu, A., 2000, ApJ 541, L54
\bibitem{}Liang, E.W., Zhang, B., O\'Brien P.T, et al., 2006, ApJ, 646, 351
\bibitem{}MacFadyen, A., Woosley, S.E., 1999, Ap. J. 524, 262
\bibitem{}Malesani, D., Tagliaferri, G.,  Chincarini, G., et al.,  2004, ApJ, 609, L5
\bibitem{}Moretti, A., Perri, M., Capalbi, M., et al. 2006, A\&A, 448, L9
\bibitem{}Nousek, J.A., Kouveliotou, C., Grupe D., et. al., 2006, ApJ, 642, 389 
\bibitem{}O' Brien, P.T., Willingale R., Osborne, J., et al., 2006, ApJ, in press astro-ph/0601125
\bibitem{}Panaitescu, A., Meszaros, P., Burrows, D., et al., 2006, MNRAS 369, 2059
\bibitem{}Panaitescu, A., astro-ph/0607396 
\bibitem{}Piro, L., De Pasquale, M., Soffitta, P., et al. 2005ApJ, 623, 314
\bibitem{}Rees, M., Meszaros, P., 1992, MNRAS, 258, 41
\bibitem{}Romano, P., Moretti, A., Banat, P. L., et al,  2006a, A\&A, 450, 59
\bibitem{}Romano, P., Campana S., Chincarini G., et al., 2006b, A\&A, in press, astro-ph/0602497. 
\bibitem{}Soderberg, A.M., Kulkarni, S.R., Nakar, E., Nature, submitted, astro-ph/0604389
\bibitem{}Stanek, K.Z., Matheson, T., Garnavich, P.M., et al. 2003, ApJ, 591, L17
\bibitem{}Thomsen, B., Hjorth, J.,  Watson, D., 2004, A\&A, 419, L21
\bibitem{}Tagliaferri, G., Goad, M., Chincarini, G., et al. 2005a, Nature, 436, 985
\bibitem{}Tagliaferri, G., Antonelli, L.A., Chincarini, G., 2005, A\&A, 443, L1
\bibitem{}Woosley, S.E., 1993, ApJ. 405, 273
\bibitem{}Woosley, S.E., Bloom, J., 2006.
\bibitem{}Woosley, S., Heger A., 2006, conf. proc. of {\it Gamma Ray Burst in the Swift Era}, astro-ph/0604131. 
\bibitem{}Woosley, S., Janka, H.T., astro-ph/0601261  
\bibitem{}Zhang, B., MacFadyen, A., Woosley, S.E., 2005, Ap.J.
\bibitem{}Zhang, B., Woosley, S.E., 2003, ASPC, 293, 321
\bibitem{}Zhang, B., Kobayashi, S., 2005, ApJ, 628, 315

\end{thebibliography}
\end{document}